\documentclass[amsmath,amssymb,twocolumn]{revtex4-1}
\usepackage{graphicx}
\usepackage{hyperref}
\usepackage{url}
\begin{document}
\title{Characterization of photoreceivers for LISA}

\author{Felipe Guzm\'an Cervantes}
\email[Corresponding author:]{felipe.guzman@nasa.gov}
\email[\\]{felipe.guzman@aei.mpg.de}
\author{Jeffrey Livas}
\author{Robert Silverberg}
\author{Ernest Buchanan}
\author{Robin Stebbins}
\affiliation{NASA Goddard Space Flight Center, Code 663, 8800 Greenbelt Road, Greenbelt, MD 20771, USA}

\begin{abstract}
LISA will use quadrant photoreceivers as front-end devices for the phasemeter measuring the motion of drag-free test masses in both angular orientation
and separation. We have set up a
laboratory testbed for the characterization of photoreceivers. Some of the limiting noise
sources have been identified and their contribution has been either measured or derived from the measured 
data. We have built a photoreceiver with a 0.5 mm diameter quadrant photodiode
with an equivalent input current noise of better than
$1.8\,\mathrm{pA}/\sqrt{\mathrm{Hz}}$ below 20\,MHz and a 3\,dB bandwidth of 34\,MHz.
\end{abstract}

\maketitle
\section{Introduction}
\begin{figure}[h]
\centering
\includegraphics[width=\columnwidth]{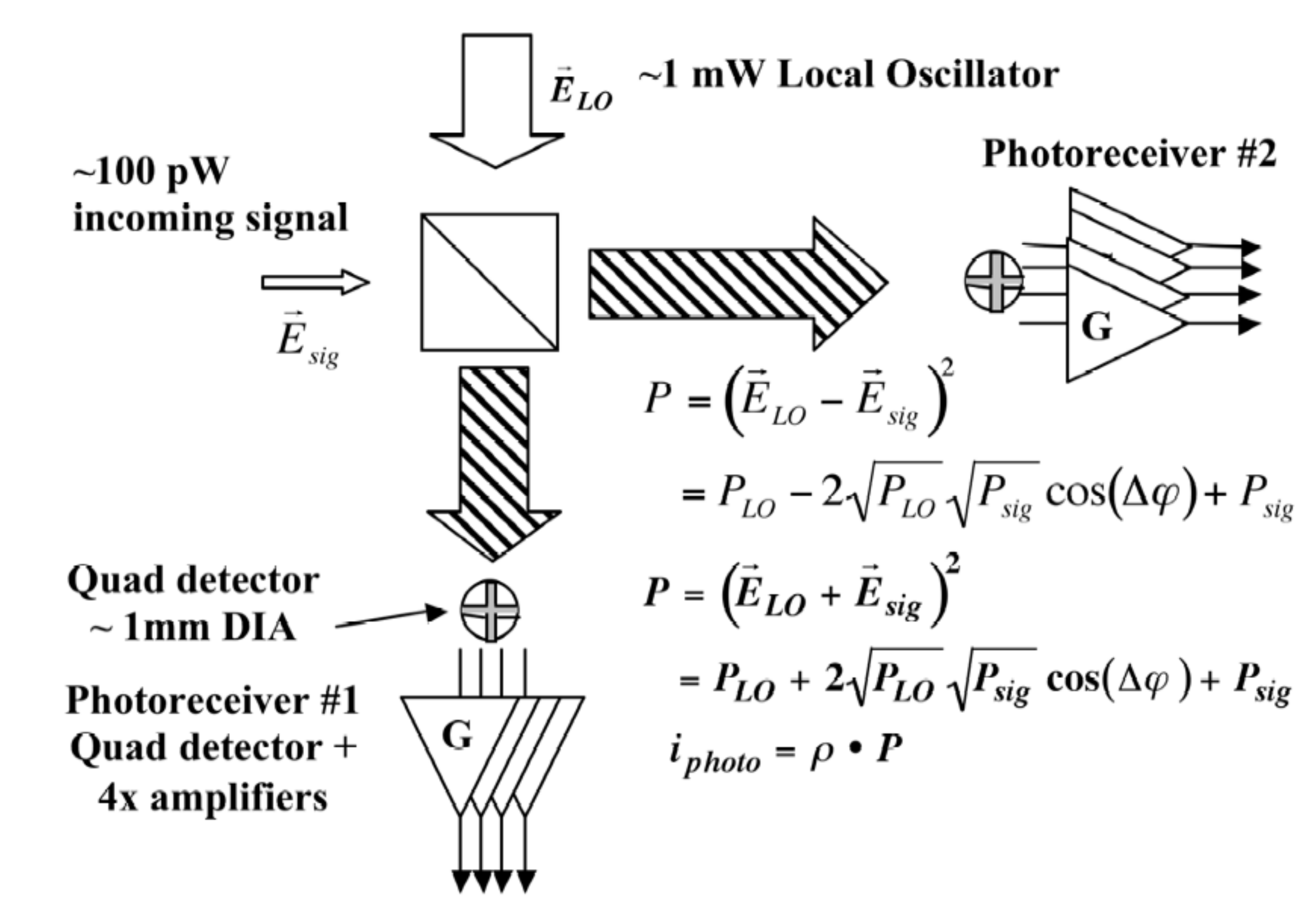}
\caption{\label{hetdetect}Descriptive diagram of laser heterodyne interferometric detection.}
\end{figure}

The Laser Interferometer Space Antenna (LISA) is a planned gravitational wave
observatory in the frequency range of 0.1\,mHz--100\,mHz that consists of three
spacecraft separated by 5 million km in a nearly equilateral triangle whose center
follows the Earth in a heliocentric orbit with an orbital phase offset of 20
degrees. Gravitational waves will be detected as distance fluctuations between
test masses moving along geodetic trajectories that are located in different
spacecraft. LISA will require low power ultra-low noise photoreceivers for
precision inter-spacecraft heterodyne laser interferometry. Quadrant
photoreceivers will be used to measure the test mass motion with a sensitivity
of 8\,$\mathrm{nrad}/\sqrt{\mathrm{Hz}}$ in angular orientation and
10\,$\mathrm{pm}/\sqrt{\mathrm{Hz}}$ in displacement
over the frequency range of 0.1\,mHz--100\,mHz\,\cite{jennrich-2009}.
The laser beam at the transmitting spacecraft will have a diameter of approximately 40\,cm and an output laser power at the telescope of the order of 1\,W.
Given the laser beam propagation over $5\times10^9$\,m and accounting for losses
on the beam path,  from the remote optical signal approximately 50\,pW will be detected on the entire quadrant photoreceiver. LISA will use heterodyne
laser interferometry (see Figure~\ref{hetdetect}) for the inter-spacecraft displacement measurement. The incoming
weak signal will optically interfere with a stronger local oscillator $P_{LO}$. The combined signal $P(t)$ measured at the photoreceiver can be expressed as
\begin{equation}
 P(t)=\underbrace{ P_{LO}+P_{sig} }_{P_{DC}:\mathrm{\,\,DC\,\,power}\,\sim\,P_{LO}}+\underbrace{2\sqrt{P_{LO}\,P_{sig}}\,\cos\left(\Delta\omega\,t+\varphi\right)}_{P_{AC}:\mathrm{\,\,heterodyne\,\,signal}},
\end{equation}
where $\Delta\omega$ is the frequency difference between the interfering laser
beams (heterodyne frequency), and $\varphi$ is the interferometer phase containing the gravitational wave information.
Both ports of the beamsplitter will be measured by quadrant detectors. Their combined information can also be used for common-mode rejection of laser
amplitude noise. The main task of the photoreceiver development is to maintain nearly shot-noise limited
performance over a measurement bandwidth from 2--20\,MHz\,\cite{shaddock-2006,bykov-2009}. This frequency range is driven by the Doppler induced frequency variations of
the optical beat note signal due to the LISA constellation armlength changes, given
by the orbits of each spacecraft.
The local oscillator power $P_{LO}$ can be adjusted, according to the required
photoreceiver performance. To reduce power consumption, temperature
gradients at the optical bench due to hot spots at the photoreceivers, and to provide additional design margin, we aim
for a low-noise wide-bandwidth photoreceiver development operating at low
$P_{LO}$ levels. Using 0.5\,mW local oscillator optical power on the entire
quadrant photoreceiver (of the order of 100\,$\mu$W per quadrant), and
assuming a responsivity $\rho$ of 0.7\,A/W for InGaAs photodiodes at a laser wavelength of 1064\,nm,
the shot-noise $i_{SN}$ can be computed as
\begin{equation}
 i_{SN}=\sqrt{2e\rho P_{DC}}\approx10\,\mathrm{pA}/\sqrt{\mathrm{Hz}}.
\end{equation}
Allocating $30\%$ of the shot-noise level --\,$3\,\mathrm{pA}/\sqrt{\mathrm{Hz}}$\,--
to the input current noise contribution of a quadrant photoreceiver, and
considering this is the quadrature sum of the current noise of the
individual quadrants, we set an input current noise
goal of $1.5\,\mathrm{pA}/\sqrt{\mathrm{Hz}}$ for the single-quadrant photoreceiver transimpedance amplifier (TIA).
\section{Photodetector transimpedance amplifier}
\begin{figure}[h]
\centering
\includegraphics[width=0.8\columnwidth]{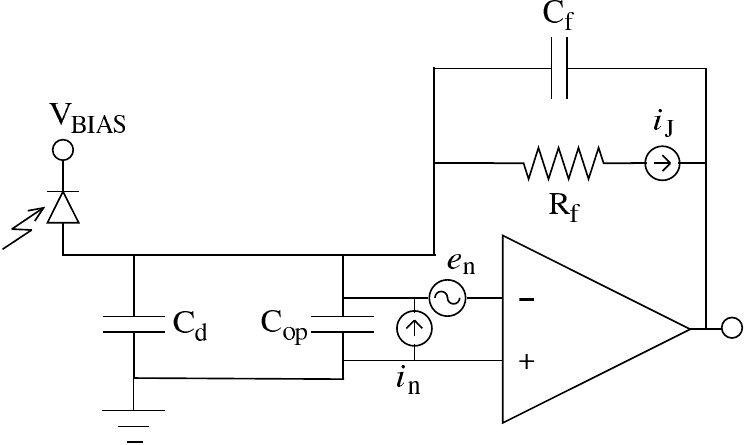}
\caption{\label{tia}Topology and noise model of photoreceiver TIA.}
\end{figure}

We have chosen a conventional DC-coupled TIA topology with a single ultra-low noise\,/\,wide-bandwidth operational amplifier (op-amp), as shown in Figure~\ref{tia}.
\subsection{Noise model}
For the TIA topology shown in Figure~\ref{tia}, two main input current noise sources have been identified in the electronics:
\begin{itemize}
 \item Johnson noise ($i_J$) from feedback resistor $R_f$:
 \begin{equation}
 i_J=\sqrt{\frac{4\,k\,T}{R_f}},
 \label{i_j}
 \end{equation}
where $k$ is the Boltzmann's constant, and $T$ is the temperature (in Kelvin).
\item Op-amp noise properties:
      The current $i_n$ and voltage $e_n$ noise properties of the op-amp contribute to the total TIA input current noise.
      \begin{itemize}
	\item The op-amp current noise $i_n$ sums directly to the TIA input.
	\item The op-amp voltage noise $e_n$ translates to current noise at the TIA input $i_{TIA}(f)$ over the input and feedback impedances as
	\begin{equation}
	  i_{TIA}(f)=e_n\frac{\sqrt{1+(2\pi\,f\,R_f\,C_{T})^{2}}}{R_f},
	\label{i_tia}
	\end{equation}
	where $f$ is the frequency, and $C_T$ is the total circuit capacitance
	\begin{equation}
	 C_T=C_d+C_f+C_{op}+C_s,
	\label{Ct}
	\end{equation}
	including the photodiode capacitance $C_d$, feedback impedance $C_f$, op-amp common-mode input capacitance $C_{op}$, and stray capacitances $C_s$ from the board, components and packaging.
       \end{itemize}
\end{itemize}
The bandwidth $BW$ of the photoreceiver can be estimated as
\begin{equation}
 BW=\sqrt{\frac{GBWP}{2\pi R_f\,C_T}},
 \label{BW}
\end{equation}
where $GBWP$ is the gain-bandwidth product of the operational amplifier.
The total TIA input current noise $I_{\mathrm{noise}}(f)$ model can be expressed as
\begin{equation}
 I_{noise}(f)=\sqrt{i_T^2+i_{TIA}^2(f)}\,\cdot\|\overline{TF}(f)\|,
 \label{i_noise}
\end{equation}
where $\|\overline{TF}(f)\|$ is the normalized TIA transfer function,
$i_{TIA}(f)$ is a frequency dependent component of the input current noise
(see Equation~\ref{i_tia}), and $i_T$ is the quadrature sum of various
contributors that can be approximated by neglecting their frequency dependency
for modeling purposes. For example, the expected current noise $i_T$ in the photoreceiver shown in Figure~\ref{tia}, can be computed as
\begin{equation}
 i_T=\sqrt{i_n^2+i_J^2+i_d^2},
 \label{it}
\end{equation}
where $i_n$ is the op-amp current noise, $i_J$ is the Johnson noise of the feedback resistor $R_f$ (see Equation~\ref{i_j}), and $i_d$ is the shot-noise from the photodiode dark current.
It can be seen from Equation~\ref{Ct} that the photodiode capacitance and the
op-amp common-mode input capacitance are crucial factors in the total noise budget. The
challenge for the photodiode manufacture lays in achieving a minimum capacitance per unit area while maintaining high responsivity and low leakage if reverse-biased. For the TIA electronics, it is necessary
to identify an op-amp with minimal common-mode input capacitance, current and voltage noise, and
a gain-bandwidth product large enough to maintain the required sensitivity over the required measurement bandwidth of 2--20\,MHz.
\section{Prototype photoreceivers}
\subsection{Collaboration with industry}
\label{DS_proto}
\begin{figure}[h]
\begin{center}
\includegraphics[width=0.8\columnwidth]{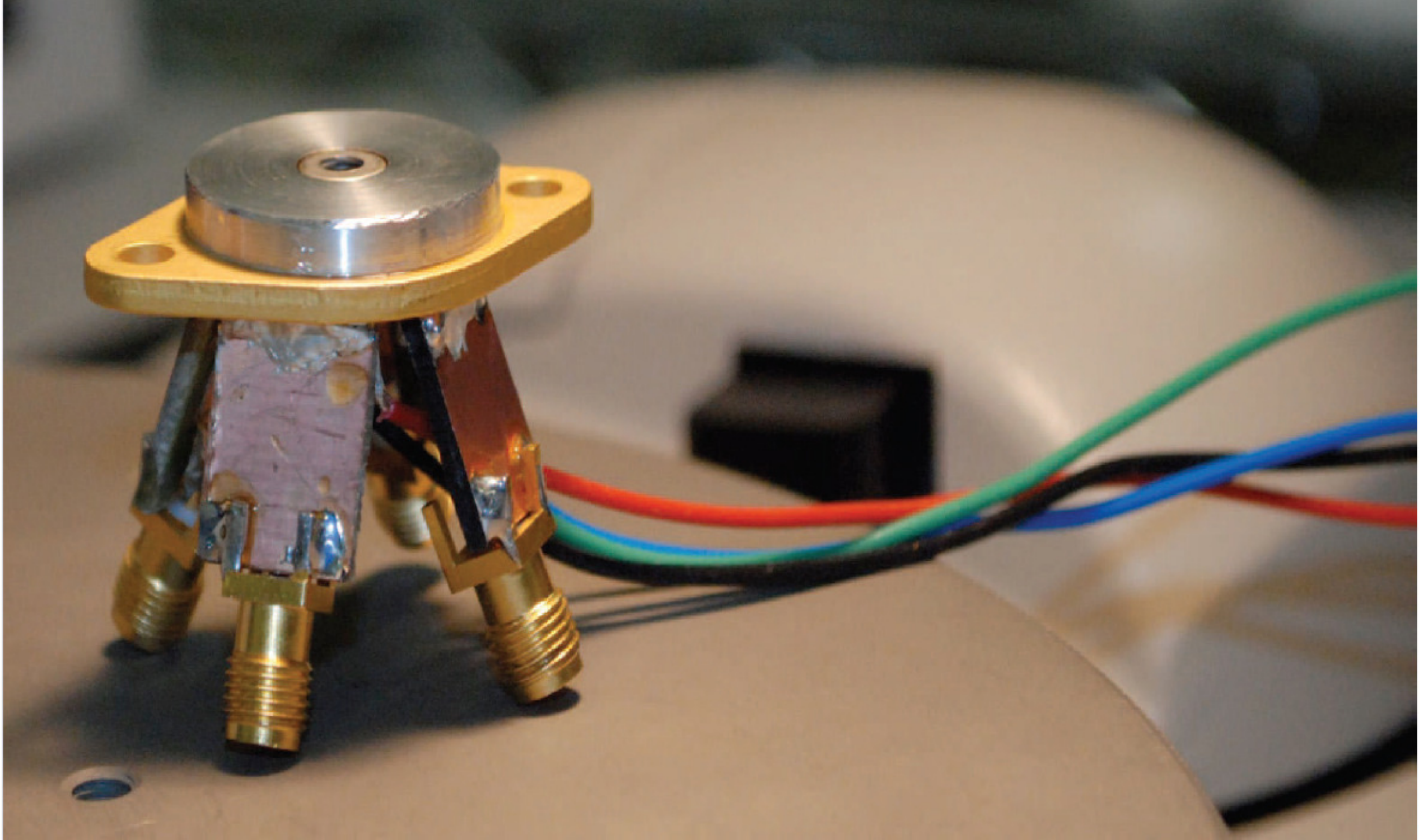}
\caption{\label{DS_quad}Photograph of a prototype quadrant photoreceiver manufactured by Discovery Semiconductors.}
\end{center}
\end{figure}

Under a Small Business Innovation Research (SBIR) grant, the company Discovery Semiconductors has developed
a large-area quadrant photodiode (QPD) of 1\,mm diameter and a quadrant capacitance of 2.5\,pF when reverse-biased at 5\,V. A
first fully integrated quadrant photoreceiver (Figure~\ref{DS_quad}: QPD + TIA electronics) performs with an equivalent input current noise of less
than 3.2\,pA/$\sqrt{\mathrm{Hz}}$ below 20\,MHz\,\cite{joshi-2009}. The characteristics of this prototype quadrant photoreceiver are:
\begin{itemize}
\item Diameter of 1\,mm with a 20\,$\mu$m inter-quadrant gap.
\item Individual quadrant capacitance $C_d$\,=\,2.5\,pF when reverse-biased at 5\,V.
\item Dark current: 140\,nA when reverse-biased at 5\,V.
\item Responsivity at 1064\,nm: $\sim0.7$\,A/W (quantum efficiency of 0.8).
\item TIA characteristics:
\begin{itemize}
\item feedback impedance: $R_f=51\,\mathrm{k\Omega}$, $C_f=0.1\,\mathrm{pF}$.
\item op-amp ADA4817: $e_n=4\,\mathrm{nV}/\sqrt{\mathrm{Hz}}$, $i_n=2.5\,\mathrm{fA}/\sqrt{\mathrm{Hz}}$, $C_{op}$\,=\,1.4\,pF, $GBWP=~410\,\mathrm{MHz}$.
\end{itemize}
\end{itemize}
Discovery Semiconductors has been awarded a second stage grant to further develop
quadrant photoreceivers. Given the successful development of a large-area low-capacitance QPD in the first step, the next stage will be focused on the noise reduction of the
electronics, e.g. by integrating op-amps with better noise properties and studying alternative TIA topologies. We
expect to receive additional devices with lower noise electronics at a later date.
\subsection{Laboratory prototypes}
\label{GSFC_proto}
Working in parallel with Discovery Semiconductors to try to understand the noise and bandwidth trade-offs in more detail, we have identified the ultra-low
noise\,/\,high-bandwidth op-amp EL5135 from Intersil\,\cite{intersil_el5135} with the following nominal noise properties:
\begin{center}
 $e_n=1.5\,\mathrm{nV}/\sqrt{\mathrm{Hz}}$, $i_n=0.9\,\mathrm{pA}/\sqrt{\mathrm{Hz}}$, $C_{op}$\,=\,1\,pF, $GBWP=1500\,\mathrm{MHz}$.
\end{center}
A significant reduction in the op-amp voltage noise together with the bandwidth enhancement (given by the higher $GBWP$) were the main factors considered
for selecting the EL5135 for laboratory prototype photoreceivers. Despite the higher op-amp current noise $i_n$ of the EL5135 compared to the
ADA4817 (mentioned above), the lower voltage noise $e_n$ of the EL5135 enables us
to achieve significantly better performance over the entire bandwidth from
2--20\,MHz. The penalty is somewhat higher noise at low frequencies. According to Equation~\ref{i_noise}, the term $i_T$
(corresponding to the quadrature sum of various current noise contributions,
including the op-amp current noise $i_n$) dominates at lower frequencies, while
the frequency dependent term $i_{TIA}$ (consisting of the op-amp voltage noise $e_n$
swing across the total equivalent TIA impedance) becomes the dominant noise
contribution at higher frequencies.\\
We have designed a TIA with a feedback impedance $R_f=40\,\mathrm{k\Omega}$, $C_f=0.1\,\mathrm{pF}$, and an expected bandwidth of
40\,MHz (according to Equation~\ref{BW}). We have built two different prototype
boards to test the noise properties of a TIA with the EL5135 op-amp:
\begin{itemize}
 \item \textbf{GAP500Q photoreceiver board}: we have chosen the commercially available QPD
GAP500Q from GPD Optoelectronics\,\cite{gpdopto_gap500q} with a diameter of
0.5\,mm and a responsivity of approximately 0.7\,A/W (quantum efficiency of 0.8) at 1064\,nm. When reverse-biased at 5\,V, this device has a nominal quadrant
capacitance $C_d$\,=\,2.0\,pF and a dark current of 2.0\,nA, according to the
manufacturer. The purpose of this investigation is to operate the TIA electronics with a photodiode that
approximates the per-quadrant capacitance and package parasitic capacitance of the larger area Discovery Semiconductors
detector.
\item \textbf{Mock-up TIA board}: we have built a board for controlled noise investigations of the TIA
performance, shown in Figure~\ref{mockup_tia}. 
\begin{figure}[h]
\centering
\includegraphics[width=0.8\columnwidth]{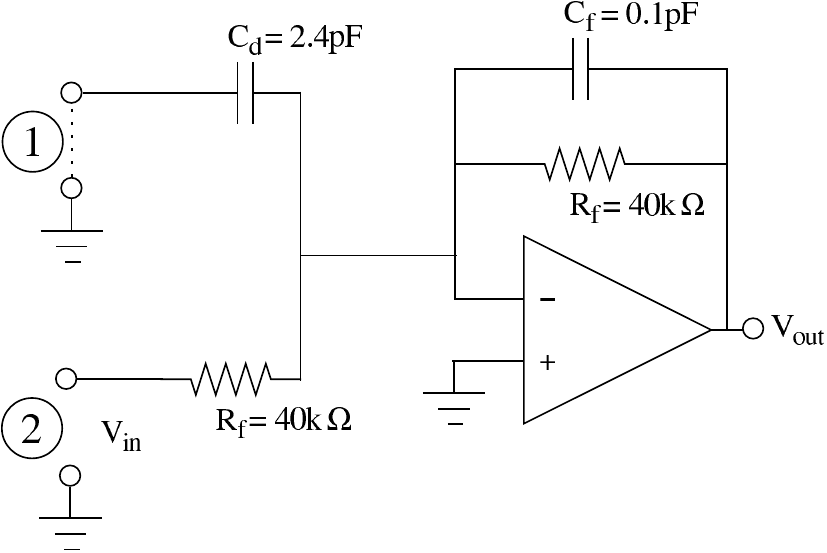}
\caption{\label{mockup_tia}Schematics of TIA mock-up board for noise measurements and frequency response measurements.}
\end{figure}

This board has two inputs:
\begin{enumerate}
 \item input 1: an input capacitor (2.4\,pF) of similar quadrant capacitance
is used to replace the photodiode. For noise measurements, this input can be grounded
while maintaining input 2 open.
\item input 2: this is used to measure the expected photoreceiver transfer
function (TF) by injecting a signal (maintaining input 1 open), and scaling it
accordingly by the feedback gain ($40\mathrm{k}$). The equivalent input current
noise can be obtained, by dividing the output voltage noise by the scaled transfer
function.
\end{enumerate}
\end{itemize}
The GAP500Q photodiode was reverse-biased with a battery power supply at 5\,V,
and the op-amps in both circuits were driven with a power supply at $\pm$5\,V.
\section{Performance measurements of prototype photoreceivers}
\begin{figure}[htb]
\begin{center}
\includegraphics[width=\columnwidth]{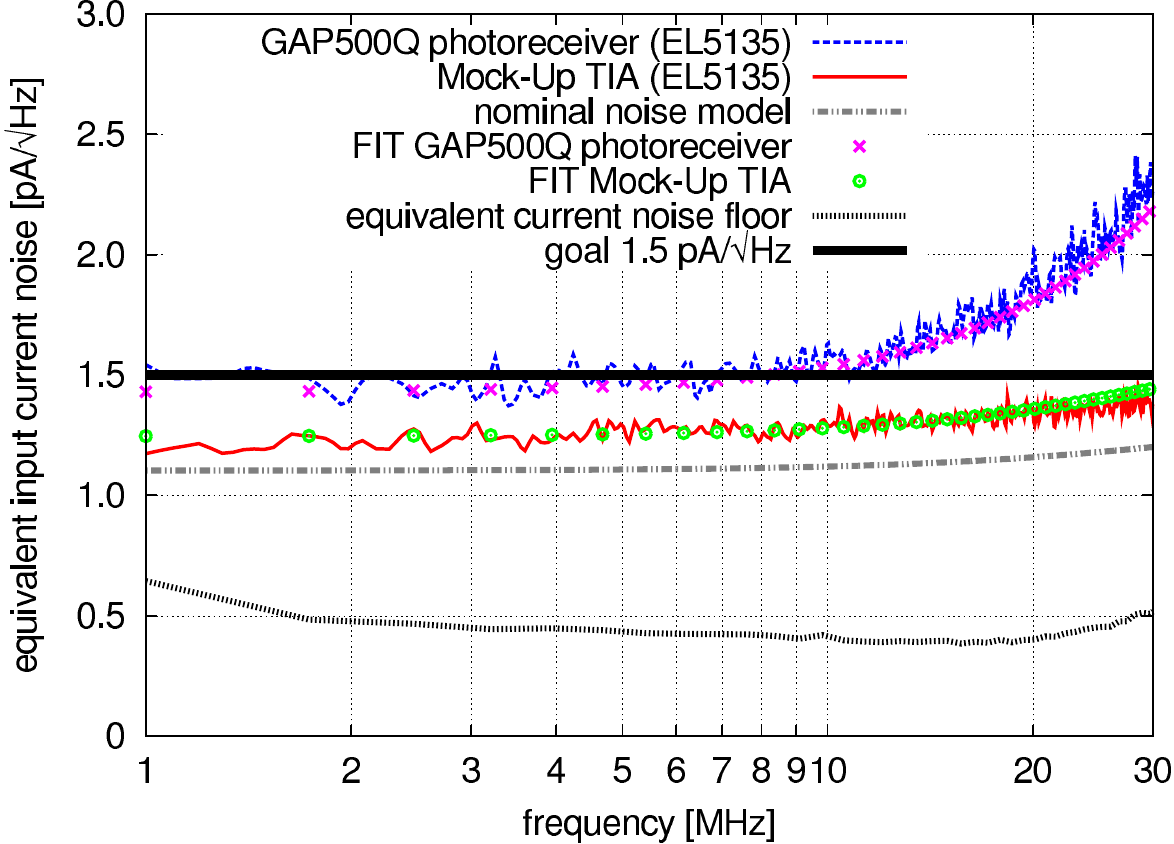}
\caption{\label{noise_meas}Input current noise measurements of photoreceiver prototypes. The
dashed trace is the photoreceiver with one quadrant of the GAP500Q QPD and the
EL5135 op-amp TIA. The solid trace is the mock-up test board (see
Figure~\ref{mockup_tia}). The dashed-dotted trace is the photoreceiver noise model for a TIA design with
an EL5135 op-amp and a quadrant capacitance of 2.5\,pF (Equations~\ref{i_noise}).  The
traces with crosses and circles show the corresponding photoreceiver noise models using parameter values 
obtained from a fit to the data. The dotted trace is the equivalent current
noise floor of the RF spectrum analyzer used as measurement instrument, referred to the input by the TIA transfer function. The
thick solid traced is our TIA input current
noise goal of $1.5\,\mathrm{pA}/\mathrm{\sqrt{Hz}}$ in the measurement band 2-20 MHz.}
\end{center}
\end{figure}

We operated the photoreceiver using only one quadrant of the GAP500Q QPD with the EL5135 op-amp for the TIA electronics.
The photoreceiver output voltage noise $V_n(f)$ is given by
\begin{equation}
V_n(f)=TF(f)\cdot\sqrt{i_{SN}^2+i_{EN}^2(f)},
\label{vn} 
\end{equation}
where $i_{SN}$ is the photocurrent shot-noise of the incident light, $TF(f)$ is
the photoreceiver transfer function, and $i_{EN}(f)$ is the equivalent input current
noise of the TIA electronics. Operating under dark
conditions, the photoreceiver output voltage noise $V_{EN}(f)$ is given by 
\begin{equation}
V_{EN}(f)=TF(f)\cdot i_{EN}(f),
\label{vdn} 
\end{equation}
By dividing Equations~\ref{vn}~and~\ref{vdn}\,\cite{diekmann-2008}, we obtain that the input current
noise of the TIA electronics can be computed as
\begin{equation}
i_{EN}(f)=\sqrt{\frac{i_{SN}^2}{\left(\frac{V_n(f)}{V_{EN}(f)}\right)^2-1}}.
\label{idn} 
\end{equation}
For equivalent input current noise measurements, we used a light-emitting diode
(LED) at a center wavelength of 1050\,nm\,($\pm$50\,nm)\,\cite{thorlabs_led1050e} as shot-noise-limited light source.
We measured the GAP500Q photoreceiver output voltage noise with ($V_n(f)$) and
without ($V_{EN}(f)$) LED light, operating at two different optical power levels
of 90\,$\mu$W and 60\,$\mu$W that are representative for the expected nominal
100\,$\mu$W per quadrant. These measurements showed
equivalent input current noise levels $i_{EN}(f)$ that did not scale with the DC
optical power level (90\,$\mu$W and 60\,$\mu$W). This also shows that the
measured current noise $i_{EN}(f)$ upon subtraction of the shot-noise
contribution, is not dependent on the optical power, which is consistent with a shot-noise behavior of the light source.\\
We also measured the output voltage noise and the transfer function ($TF(f)$) of our mock-up TIA
board (see Figure~\ref{mockup_tia}). Analogous to Equation~\ref{vdn}, the
input current noise can be computed by referring the output voltage noise (upon
subtraction in quadrature of the RF spectrum analyzer voltage noise floor) to the input
dividing by the transfer function. Figure~\ref{noise_meas} shows the noise measurements.
The GAP500Q photoreceiver reaches a level of about
$1.5\,\mathrm{pA}/\mathrm{\sqrt{Hz}}$ up to $\sim$10\,MHz, increasing at
higher frequencies. It exceeds the noise goal by approximately 20\%
($1.8\,\mathrm{pA}/\mathrm{\sqrt{Hz}}$) at 20\,MHz. The mock-up TIA circuit
meets our noise goal over the entire bandwidth (2-20\,MHz), however, it shows noise in excess
of the model. At lower frequencies the equivalent input noise
is determined by excess current noise ($i_T$: assumed to have no frequency
dependence, Equation~\ref{it}), while at higher frequencies, the increasing
slope is dominated by the op-amp voltage noise swing across the total circuit
capacitance ($i_{TIA}$: dependent on frequency, Equation~\ref{i_tia}).
We run a set of measurements in order to determine some of the involved unknowns:
\begin{enumerate}
\item op-amp voltage noise ($e_n$): we measured the op-amp voltage noise on a separate sample of the
EL5135, by driving the op-amp as a voltage follower with grounded input.
A low-noise 10x amplifier in series with the op-amp output was necessary for a voltage noise
measurement above the spectrum analyzer noise floor. The voltage noise level
measured was $e_n=2.1\,\mathrm{nV}/\mathrm{\sqrt{Hz}}$ at 10\,MHz, which is
significantly higher than the specified $1.5\,\mathrm{nV}/\mathrm{\sqrt{Hz}}$.
\item photodiode and feedback capacitances ($C_d$, $C_f$): using a LCR-bridge impedance measurement instrument, we measured the photodiode quadrant capacitance $C_d$ on
a separate sample of the GAP500Q to be 3.2\,pF with a 5\,V reversed bias, which
is higher than the nominal 2\,pF. We also measured the capacitor $C_d$ of the
mock-up board and the feedback capacitor $C_f$ to be 2.2\,pF (nominally 2.4\,pF) and 0.1\,pF,
respectively.
\end{enumerate}
These noise properties are higher than specified and may account for part of the excess
noise. A direct measurement of the op-amp current noise $i_n$, the
op-amp input capacitance $C_{op}$, and the stray capacitances $C_{s}$ of the circuit
involve the development of dedicated electronic boards (currently on-going) for well-controlled measurements. These measurements will be conducted at a later stage. 
However, it is possible to obtain an estimate of these values by fitting them as
parameters of the noise model (Equations~\ref{i_noise}~and~\ref{it}) to the two measured data sets. The noise level difference
(offset) in the data of the GAP500Q photoreceiver and the mock-up TIA is an indicator
of a significant excess current noise contribution, $i_X$, present in the
photo-measurement and not in the measurement of the mock-up TIA. This can
also be included into the fit by considering the following non-frequency dependent contributions $i_T$ (Equation~\ref{it}) for each case:
\begin{eqnarray}
\mathrm{GAP500Q\,\,photoreceiver:} &  \nonumber\\
 i_{T_{PD}} & =  \sqrt{i_n^2+i_J^2+i_d^2+i_X^2},\\
\mathrm{Mock-up\,\,TIA:} &  \nonumber\\
 i_{T_{MU}} & =  \sqrt{i_n^2+i_J^2}.
\label{i_Ts}
\end{eqnarray}
The capacitance values $C_{d}$ and $C_{f}$ are assumed to be known from the LCR-bridge
measurements. We also assume similar op-amp and board noise properties ($e_n$,
$i_n$, $C_{op}+C_{s}$) for the two circuits. From the fit, we obtained an op-amp current noise of
$i_n\approx1.1\,\mathrm{pA}/\mathrm{\sqrt{Hz}}$ (about 20\% higher than nominal $0.9\,\mathrm{pA}/\mathrm{\sqrt{Hz}}$) and a combined stray plus op-amp
input capacitance $C_{op}+C_{s}\approx1.3\,$pF (1\,pF nominal $C_{op}$). We also fit a common op-amp voltage
noise for the two data sets to be
$e_n\approx1.9\,\mathrm{nV}/\mathrm{\sqrt{Hz}}$, which is comparable (within
$<10\%$) to the independent measurement ($2.1\,\mathrm{nV}/\mathrm{\sqrt{Hz}}$), but about 25\% higher than nominal ($1.5\,\mathrm{nV}/\mathrm{\sqrt{Hz}}$).
Table~\ref{fitvalues} summarizes the current best estimates (CBE) of the photoreceiver noise properties.
\begin{table}[ht]
\centering
\begin{tabular}{c | c  c c}
\\
parameter & nominal & CBE & method  \\
\hline\\
$C_d\,\left[\mathrm{pF}\right]$ & 2.0 & 3.2 & measured \\ \\
$e_n\,\left[\mathrm{nV}/\mathrm{\sqrt{Hz}}\right]$ & 1.5 & 1.9 & fit \\ \\
$i_n\,\left[\mathrm{pA}/\mathrm{\sqrt{Hz}}\right]$ & 0.9 & 1.1 & fit \\ \\
$C_{op}+C_s\,\left[\mathrm{pF}\right]$ & 1.0 & 1.3 & fit \\ \\
$i_X\,\left[\mathrm{pA}/\mathrm{\sqrt{Hz}}\right]$ & - & 0.7 & fit \\ \\
\end{tabular}
\caption{\label{fitvalues}Noise parameters: comparison between nominal values and current best estimates (CBE).}
\end{table}The excess current noise contribution, $i_X$, present in the GAP500Q
photoreceiver data was determined to be of the order of
$i_X\approx0.7\,\mathrm{pA}/\mathrm{\sqrt{Hz}}$. We have reversed-biased the
photodiode with a battery power supply, therefore, noise on the bias voltage translating
to excess current noise is not the cause. Additional tests are required to determine the
origin of this contribution.
\section{Conclusions and Outlook}
We have presented the results of noise measurements conducted on different
photoreceiver prototypes. The measurements showed approximately 20\% noise in
excess of our goal between 10--20\,MHz. Direct measurements of the
op-amp voltage noise and the reverse-biased QPD quadrant capacitance evidenced noise levels higher than nominal, accounting for part
of the excess noise at higher frequencies. By fitting the parameters of the
noise model to the data, we obtained estimates for the combined stray plus op-amp
input capacitance and the op-amp current noise $i_n$, which was determined to be
approximately 20\% higher than nominal. Significant excess current noise (50\% of total) $i_X$
was determined between photoconductive (GAP500Q photoreceiver) and
electronic (mock-up TIA) noise measurements. Additional testing is required to
determine its origin. The measured photoreceiver performance is of the order of
$1.5\,\mathrm{pA}/\mathrm{\sqrt{Hz}}$ below 10\,MHz, increasing up to
$1.8\,\mathrm{pA}/\mathrm{\sqrt{Hz}}$ at 20\,MHz with a 3\,dB bandwidth of 34\,MHz. However, the mock-up TIA
performs at a level of $1.35\,\mathrm{pA}/\mathrm{\sqrt{Hz}}$ below 20\,MHz
(10\% higher than expected from the nominal model) with a measured 3\,dB
bandwidth of 38\,MHz. This suggests a significantly better performance of a real
photoreceiver with the current TIA design, depending upon clarification and, if viable, mitigation of the excess current noise $i_X$.
In addition, as following steps, we plan to conduct spatial scanning of the
photodiode surfaces, measurement of inter-quadrant cross-talk, and differential wavefront sensing angle measurements.
\section{Acknowledgements}
This research was supported in part by NASA contract ATFP07-0127. F. Guzm\'an Cervantes is supported by an appointment to the NASA Postdoctoral Program
at the Goddard Space Flight Center, administered by Oak Ridge Associated Universities (ORAU) through a contract with NASA. We thank A. Joshi, S. Datta and J. Rue for stimulating discussions.
%

\end{document}